\documentclass[twocolumn,showpacs,preprintnumbers,amsmath,amssymb,prb,floatfix]{revtex4}

\usepackage{graphicx}% Include figure files
\usepackage{dcolumn}% Align table columns on decimal point        
\usepackage{bm}% bold math
\usepackage{mhchem} 
\usepackage{epstopdf}     

\begin{document}

\title{Gutzwiller Variational Method for Intersite Coulomb Interactions: The Spinless Fermion Model in One Dimension}       

\author{M. Sherafati}
\author{S. Satpathy}

\affiliation{
Department of Physics $\&$ Astronomy, 
University of Missouri,    
Columbia, MO 65211}

\author{D. Pettey}

\affiliation{
Department of Mathematics, 
University of Missouri,     
Columbia, MO 65211}

\date{\today}     

\begin{abstract}
We study the Gutzwiller method for the spinless fermion model in one dimension, which is one of the simplest models that incorporates the intersite Coulomb interaction. The Gutzwiller solution of this model has been studied in the literature but with differing results. We obtain  the Gutzwiller solution of the problem by a careful enumeration of the many-particle configurations
and explain the origin of the discrepancy in the literature to be due to the neglect of the correlation between the neighboring bond occupancies.  
The correct implementation of the 
Gutzwiller approach yields results different from the slave-boson mean-field theory, unlike for the Hubbard model with on-site interaction, where both methods are known to be equivalent. The slave-boson    and the Gutzwiller results are compared to the exact solution, available for the half-filled case, and to the numerical exact diagonalization results for the case of general filling. 
%Interestingly, the model indicates phase separation behavior for sufficiently large band filling, which is faithfully reproduced by the Gutzwiller solution as well.
\end{abstract}
\maketitle 

%: Introduction

{\it Introduction} --  Conventional many-body theory has difficulty in taking into account the local correlations while, at the same time, preserving the itinerant character of the many-electron states. Because of this, there has been a resurgence of interest in the variational Gutzwiller method,\cite{Gutzwiller} where local correlations show up as a renormalization of the kinetic energy\cite{Gutzwiller-ReductionFactor, Ogawa, Vollhardt} within the Gutzwiller approximation  and the overall itinerant character of the electrons is preserved.  The method leads to a many-body modification of the Fermi surfaces and is being increasingly used to describe the measured Fermi surface topology of correlated systems starting from materials as simple as   Ni\cite{Bunemann-Ni} to more complex systems such as NaCoO$_2$\cite{Zhou}  and  LaOFeAs\cite{Andersen} using model Hamiltonians. This in turn has led to a flurry of work for incorporating the   Gutzwiller approach in the density-functional theory for realistic solids.\cite{Julien2006,GDFT1,GDFT2} 

While the Gutzwiller method has been well developed for on-site Coulomb interactions, it has been less studied when the inter-site Coulomb interactions are present. On the other hand, the nearest-neighbor  Coulomb interaction is an important ingredient in models used to describe the charge order-disorder transition in many systems such as  the Verwey transition in the half-metallic Fe$_3$O$_4$. Thus there is a need to develop the Gutzwiller approach for incorporating the effects of the  inter-site Coulomb interactions, which presents nuances not encounterd in treating models with on-site Coulomb interactions such as the Hubbard model. 

The simplest model that contains the intersite Coulomb interaction is the spinless fermion model  
 in one dimension (1D)
 \begin{equation}
 {\hat H} = -t \sum_{ij} \hat c_i^\dagger \hat c_j + h. c. + V \sum_{ij} \hat n_i \hat n_j,  
 \label{Hamil}
 \end {equation}
where $\hat c_i^\dagger$ creates a spinless fermion at the site $i$, $ \hat n_i = \hat c_i^\dagger \hat c_i$, and the summation is taken over distinct pairs of nearest neighbors. 
The is also an important model due to its equivalence to the XXZ Heisenberg model via the Jordan-Wigner transformation.\cite{Giamarchi}
 The Gutzwiller approach for this model has been treated by three different papers in the literature\cite{Fazekas1, Fazekas2, Seibold}  including two by the same author; however the results differ from one another including the all-important Gutzwiller reduction factor for the kinetic energy.  In this paper, we study this model with the Gutzwiller approach by a careful enumeration of the many-particle configurations. 
 Our results agree with one\cite{Fazekas2} of the two papers of Fazekas, while we find that the reason for disagreement with the third paper\cite{Seibold} is their neglect of the correlation between the states of the nearest-neighbor bonds 
 %(we use the term ``bond" to describe a pair of nearest-neighbor sites) 
in counting the many-particle configurations. This omission of the correlation, both in the Gutzwiller  as well as in the slave-boson mean-field methods, effectively turns the bond problem into the site problem,\cite{Seibold} yielding results for the 1D spinless fermion model very similar to the 1D Hubbard model. 
%Apart from resolving the discrepancy in the literature, we present the results for the case of general filling and show that for relatively higher band filling, the homogeneous phase is unstable towards the formation of a mixed phase.

%: Gutzwiller Psi_G
{\it Gutzwiller wave function} --
The Gutzwiller variational wave function is written as
\begin{align}
|\Psi_G \rangle =g^{\hat B} |\Psi_0 \rangle,
\label{Gutz}
\end{align}
where $|\Psi_0 \rangle$ is the uncorrelated wave function, $g$ is a variational parameter between 0 and 1, and
the Gutzwiller factor $g^{\hat B} $,  with the bond occupancy operator $\hat  B = \sum \hat  n_i \hat n_{i+1}$, reduces the weight of the configurations where neighboring sites are occupied ($B=1$). 
The uncorrelated wave function $|\Psi_0 \rangle$  can itself be taken as variational as well.\cite{Julien2006}
The variational parameter $g$  is determined by minimizing the total energy $ \langle {\cal \hat H} \rangle_G =  \langle \Psi_G | {\cal \hat H}   |\Psi_G \rangle /  \langle \Psi_G   |\Psi_G \rangle  $, 
which  however is difficult to evaluate analytically for any given $ |\Psi_0 \rangle$ except for special cases.
One then resorts to the Gutzwiller approximation for the expectation values of the one-particle operators: 
$ \langle \hat{O} \rangle_G \equiv  \langle \Psi_G | \hat{O}   |\Psi_G \rangle /  \langle \Psi_G   |\Psi_G \rangle  \ $
$\approx  \gamma  \langle \Psi_0 | \hat{O}   |\Psi_0 \rangle /  \langle \Psi_0   |\Psi_0 \rangle$,
where the ``reduction factor" $\gamma$  is independent of the $| \Psi_0\rangle$ used in the Gutzwiller wave function. 
This is the Gutzwiller approximation. However, if one  uses a  $| \Psi_0\rangle$ for which $\gamma$ is exactly evaluated, then the results are fully variational as is the case here.

%:Equal-weight Psi
Gutzwiller's original approach for computation of the reduction factor turns out to be\cite{Fulde} exact if one uses the
equal-weight many-body configuration for the uncorrelated state:  
\begin{equation}
|\Psi_0 \rangle   = N_\Gamma^{-1/2}  \sum |\Gamma \rangle,
\label{Psi-equal}
\end{equation}
in which case the calculation of $\gamma$ boils down to merely a combinatorial problem. Here $N_\Gamma$ is the total number of many-body configurations. 
%Our treatment follows closely the case of the Hubbard model with the on-site interaction, where the site double occupancy $\hat D$ appears in the Gutzwiller factor instead of the bond occupancy factor in Eq.\   \eqref{Gutz}. 
The essential complication here is that while the site occupancies in the Hubbard model are specified independently of one another, the bond occupancies are correlated between nearest-neighbor bonds, making the combinatorial problem significantly more involved. 
%In fact, we have not been able to solve the combinatorial problem  in higher dimensions or even for the simpler, Bethe lattice case.

Note that, as already mentioned. since we evaluate  the reduction factor  for the equal-weight state (\ref{Psi-equal}) exactly and we also use the same state in the Gutzwiller variational wave function (\ref{Gutz}) throughout this paper, our results remain fully variational. We have not tested a better variational wave function for $| \Psi_0 \rangle$   in  Eq. (\ref{Gutz}), since our main goal here is to obtain the Gutzwiller reduction factor.
%
%: Fig 1
\begin{figure} [t]
\includegraphics[angle=0,width=0.9   \linewidth]{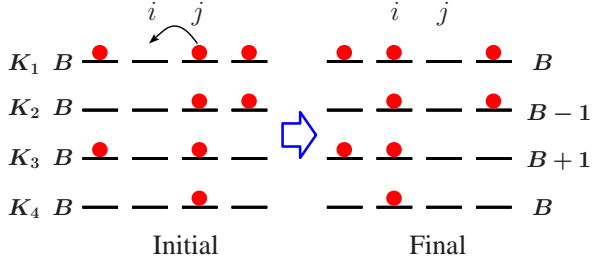}
\caption{(Color online)   Initial and final configurations, $\Gamma_\mu$ and $\Gamma^\prime_\mu$, $\mu = 1, ..., 4$, 
that contribute to the kinetic energy term $\langle c_i^\dagger c_j \rangle$. The change in the total number of bonds between the initial and the final configurations are indicated as well as the total number of configurations $K_\mu (B, N, L)$ in each case.
}
\label{config}
\end{figure}

Proceeding now with the evaluation of the expectation values, we start with the norm
\begin{equation}
\langle \Psi_G | \Psi_G \rangle      = N^{-1}_\Gamma   \sum^{N-1}_{B=0} g^{2B} K(B,N,L), 
 \label{eq:PsiGBnorm}
\end{equation}
where $K(B,N,L)$, evaluated explicitly in Appendix I,
 is the number of configurations   with the bond occupancy $B$, $L$ is the number of sites on the 1D lattice (periodic boundary condition is assumed, so that we have a ring instead of a chain), and $N (\le L)$ is the number of electrons. In the thermodynamic limit, $ N , L, B \rightarrow \infty$ and $n = N/L$ is finite, the summand is a sharply peaked function of $B$, and the sum may be replaced by the largest value determined from the condition
$\partial (g^{2B} K(B,N,L) ) / \partial B = 0$.  This yields the relation between the variational parameter $g$ and the bond occupancy per site $b = B/L$, which reads
\begin{equation}
g^2 =\frac{b(1-2n+b)}{(n-b)^2},
\label{g2}
\end{equation}
or inverting this, we find
\begin{equation}
b =n + \frac{p}{2} (g^2-1)^{-1},
\label{b}
\end{equation} 
where 
\begin{equation}
 p \equiv  1- [1+4n(g^2-1) (1-n)]^{1/2}.
\label{p}
\end{equation} 
It can be easily verified that for the uncorrelated wave function, $g =1$, we recover the correct result $\lim_{g\rightarrow 1} b = n^2$. The final result for the norm is 
\begin{equation}
\langle \Psi_G | \Psi_G \rangle      =      \big[ \frac{b(1-2n+b)}{(n-b)^2}   \big]^B \times    \frac{  K(B,N,L) } { N_\Gamma }. 
\label{Norm} 
\end{equation}

In view of the fact that in the thermodynamic limit, $|\Psi_G\rangle$ is an eigenstate of the bond occupancy operator $\hat B$, the total energy can be evaluated as
\begin{align}
E(B) &= \frac {\langle \Psi_G | \hat H |\Psi_G \rangle} {\langle \Psi_G |\Psi_G \rangle}  \nonumber \\ 
&= -t \sum_{ij} \gamma  \frac { \langle \Psi_0 |   \hat c_i^\dagger \hat c_j |\Psi_0\rangle  } 
{\langle \Psi_0 |  \Psi_0\rangle}  + h. c. + VB,
\label{EB}
\end{align}
where $\gamma \equiv \langle    \hat c_i^\dagger \hat c_j \rangle_G / \langle    \hat c_i^\dagger \hat c_j \rangle_0$ is the reduction factor of the 
kinetic energy due to correlation effects.

% 
%: Fig 2 
\begin{figure} 
\includegraphics[angle=0,width=0.8   \linewidth]{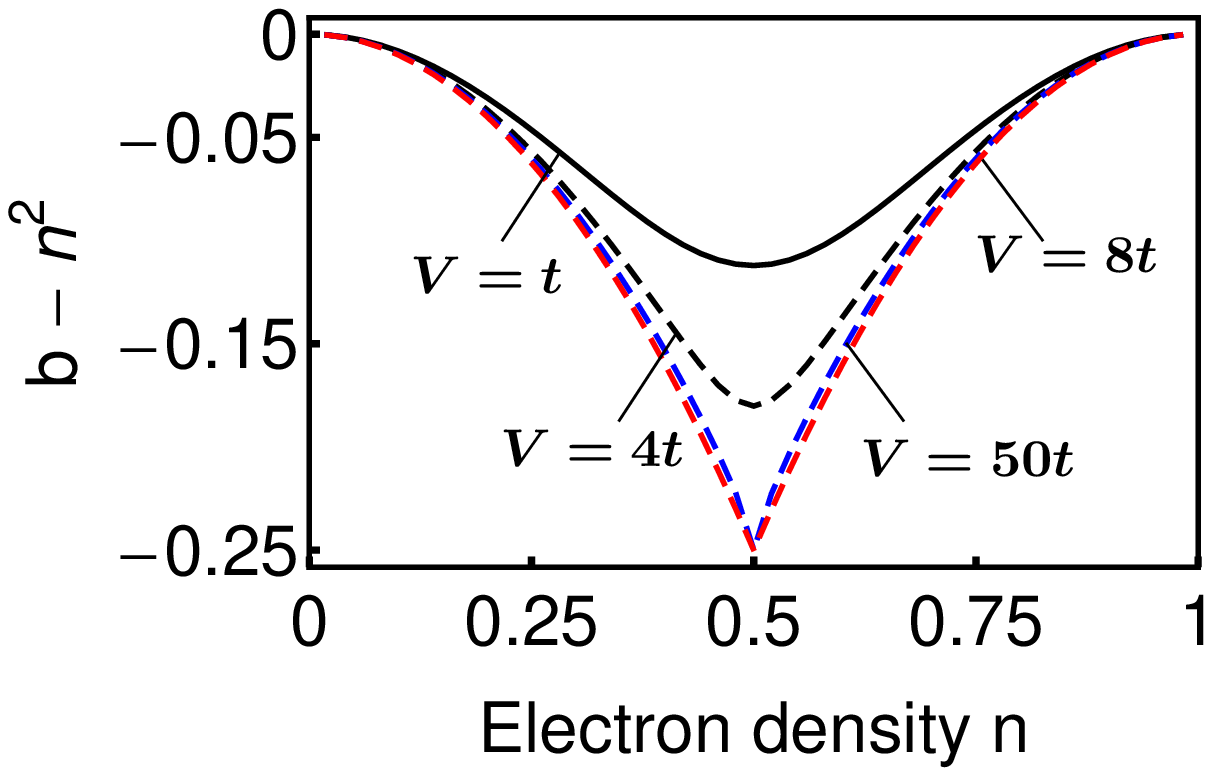}
\includegraphics[angle=0,width=0.8   \linewidth]{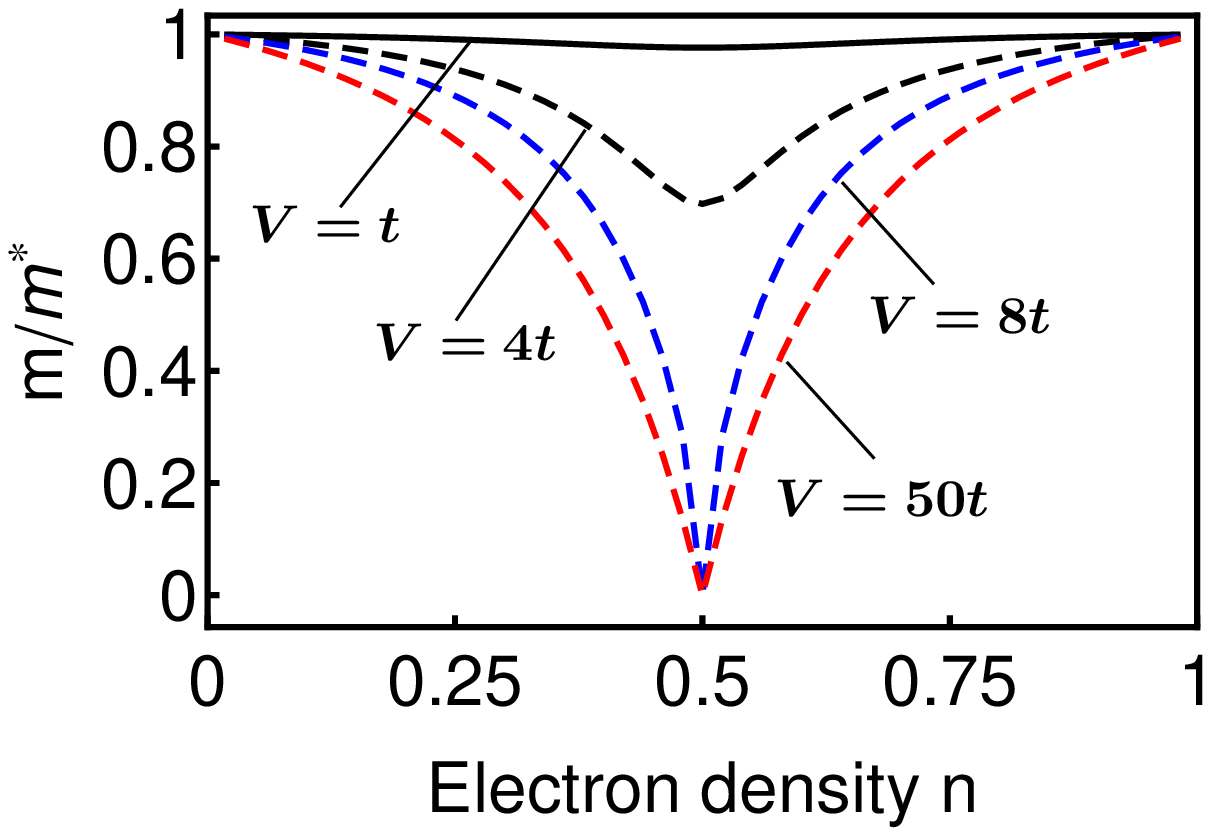}
\caption{(Color online)    Reduction of the bond occupancy $b-n^2$  from the uncorrelated value ($b_{\text {uncorr}} = n^2$)  ({\it top}) and the  inverse renormalized band mass $m/m^*$ ({\it bottom})  as a function of the band filling $n$ for several values of the interaction
$V$. For half filling, $n=1/2$, a metal-insulator transition occurs for $ V \ge 8t$ within the Gutzwiller method, where the bond occupancy becomes zero and the renormalized mass becomes infinity.
}
\label{bgamma} 
\end{figure}

{\it Gutzwiller reduction factor $\gamma$} -- 
To obtain the expression for $\gamma$, we compute the expectation value
\begin{align}
\langle  \Psi_G | &   \hat c_i^\dagger \hat c_j |\Psi_G \rangle = \frac{1}{N_\Gamma} \sum_{R, \mu} \langle \Gamma^\prime_\mu \Gamma_R| g^{\hat B} 
\hat c_i^\dagger \hat c_j         g^{\hat B}                    |  \Gamma_\mu \Gamma_R \rangle \nonumber \\
&=  \frac{1}{N_\Gamma} \sum_B g^{2B} (K_1 +g^{-1} K_2 + g K_3 + K_4),  
\label{GTG}
\end{align}
where in the configuration $| \Gamma \rangle \equiv | \Gamma_\mu \Gamma_R \rangle $, $\Gamma_\mu$ refers to the configuration of the four neighboring sites shown in Fig.\   \ref{config}, while $\Gamma_R$ refers to the remaining sites. Only the four initial and final configurations contribute to the sum and the $K_i$'s are simply the total number of configurations $\Gamma_R$ in each of the four cases
shown in the figure. Using again the thermodynamic limit to replace the sum in the second line of Eq.\   \eqref{GTG} by its largest value and dividing by  the norm Eq.\   \eqref{Norm}, we find
\begin{align}
&\langle    \hat c_i^\dagger \hat c_j \rangle_G 
\equiv \frac{\langle  \Psi_G |    \hat c_i^\dagger \hat c_j |\Psi_G \rangle}  {\langle  \Psi_G |  \Psi_G \rangle}
 = \frac{K_1 +g^{-1} K_2 + g K_3 + K_4}{K}   \nonumber \\
 & =\frac {(b-bg+gn) (1-2n + b-bg+gn) (n-b)}{gn (1-n)},
 \label{GTG2}
\end{align}
where the final expression has been obtained by using the results given in the Appendix for the $K$'s.  
 By putting $g=1$ for the uncorrelated state, we readily
get $\langle    \hat c_i^\dagger \hat c_j \rangle_0 = n(1-n)$
and
after some algebra  an expression for the all-important kinetic energy reduction factor in the Gutzwiller theory is found. It takes a much more complex form than the corresponding factor for the Hubbard model with the on-site Coulomb interaction, viz.,
%
%\begin{align}
%\gamma =  \frac { \langle    \hat c_i^\dagger \hat c_j \rangle_G }{ \langle    \hat c_i^\dagger \hat c_j \rangle_0} = 
%\frac { p [p-2n (1+g)] [p+2 (n-1) (1+g)]}       {  8n^2 (1-n)^2g (1-g) (1+g)^3 },
%\label{gamma}
%\end{align}
%
\begin{align}
\gamma =  \frac { \langle    \hat c_i^\dagger \hat c_j \rangle_G }{ \langle    \hat c_i^\dagger \hat c_j \rangle_0} = 
\frac { p - 2n(1-n) (1+g) (p+1-g)    }    
{  2n^2 (1-n)^2       (g-1) (1+g)^3 },
\label{gamma}
\end{align}
which is expressed in terms of the single variational parameter $g$.  

The expression for the reduction factor, Eq. (\ref{gamma}),  is the central result of this work and it is valid for any general filling factor $n$.
Three mutually conflicting expressions were obtained in the literature earlier\cite{Fazekas1, Fazekas2, Seibold} for this reduction factor. We find that our expression agrees with the second paper by Fazekas\cite{Fazekas2}, which has the expression for $\gamma$ for the half-filled case only. As explained in the section on the slave-boson solution below,  the origin of the discrepancy in the literature is due to the neglect of the correlation between the neighboring bond occupancies, so that the reduction factor in Ref. \onlinecite{Seibold}, e.\ g., becomes identical to the case for the on-site Hubbard model.

%: Fig 3 
\begin{figure} [h]
\includegraphics[angle=0,width=0.85   \linewidth]{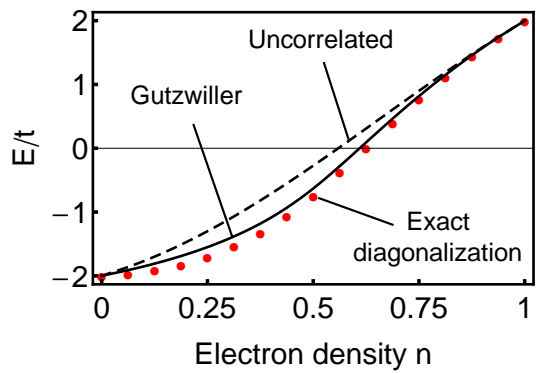}
\caption{(Color online)    Ground-state energy  per electron as a function of electron density  calculated with the correlated (solid line) and uncorrelated (dashed black line) wave functions. The red dots indicate the results obtained from exact diagonalization on a 16-site lattice with periodic boundary condition.  
%The model shows phase separation with the dashed blue  line showing the energy of the system in the phase-separated region. 
The Coulomb interaction parameter used here is  $V/ t =4$. 
}
\label{Eperelectron}
\end{figure}

{\it Total energy} -- The total energy $E(g, n)$ may be readily written using the Eqs.\   \eqref{EB},   \eqref{GTG2}, and   \eqref{b} for any band filling  $n$ (which is the same as the electron density).  The ground-state energy is obtained from the condition $\partial E / \partial g = 0$. The results for the band occupancy $b$ and the renormalized mass $m^*/m = \gamma_{\text {uncorr}} / \gamma$ are shown in Fig.\   \ref{bgamma}, where a Brinkman-Rice\cite{Brinkman}  type metal-insulator transition is indicated for the half-filling ($n = 1/2$) beyond the critical value of the interaction, $ V \ge 8t$. At the Brinkman-Rice transition point, the effective mass $m^*$ goes to infinity and the band occupancy goes to zero, as seen from Fig. \ref{bgamma}.

\begin{figure} 
\includegraphics[angle=0,width=0.85   \linewidth]{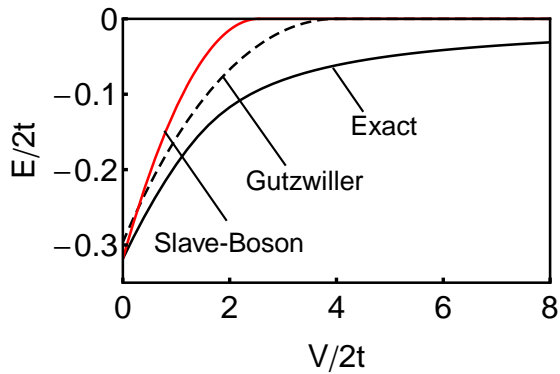} 
\caption{(Color online)   Exact ground-state energy per site for the spinless fermion Hamiltonian compared with the same obtained from the  Gutzwiller and the slave-boson mean-field methods. 
}
\label{GS-energy}
\end{figure}

The calculated total energy as a function of the band filling is shown in Fig.\ \ref{Eperelectron}, where it is also compared to the exact results for the 16-site lattice obtained from the exact diagonalization using the Lanczos method. 
Note that the Gutzwiller energy forms an upper bound to the exact energy here, since with the equal-weight state as the uncorrelated state for 
$|\Psi_o \rangle $ in Eq.\ \eqref{Gutz}, the method is fully variational without any approximation, because all expectation values are evaluated exactly in this case. Only when we use a different uncorrelated state for $|\Psi_o \rangle $, but at the same time still use the expression for the reduction factor $\gamma$ obtained using  the equal-weight state (Gutzwiller approximation), that the variational principle does not hold any more. Therefore, in both the Figs.\  \ref{Eperelectron} and \ref{GS-energy}, the Gutzwiller energy forms a strict variational upper bound to the exact energies.

\begin{figure} [t]
\includegraphics[angle=0,width=0.7   \linewidth]{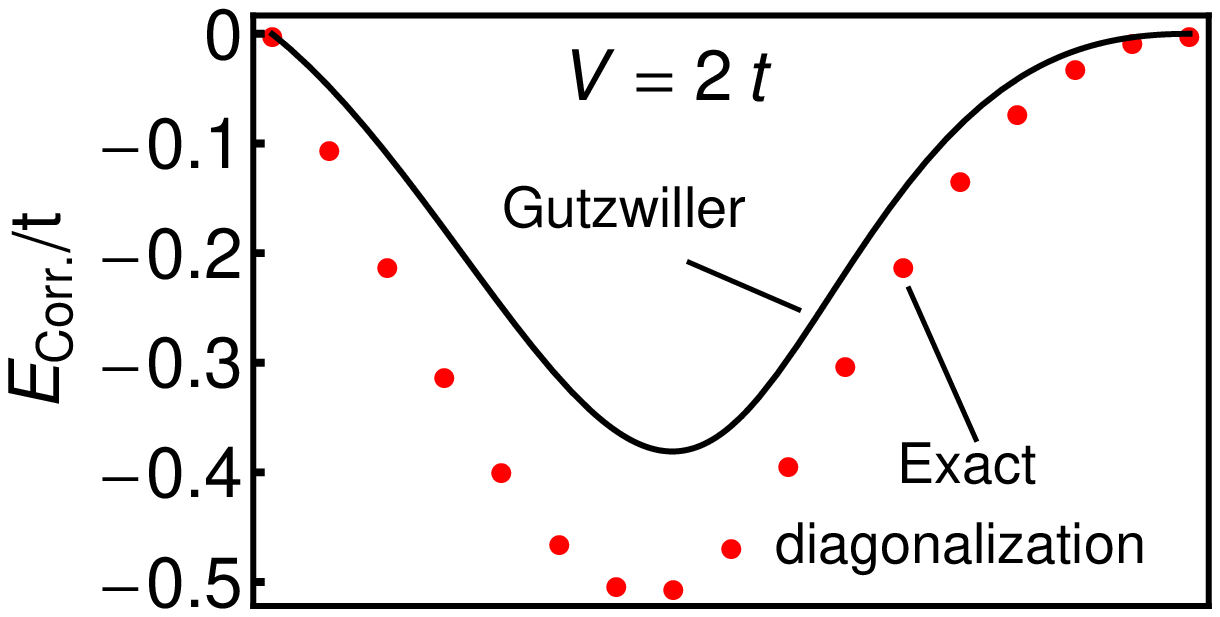}
\includegraphics[angle=0,width=0.7   \linewidth]{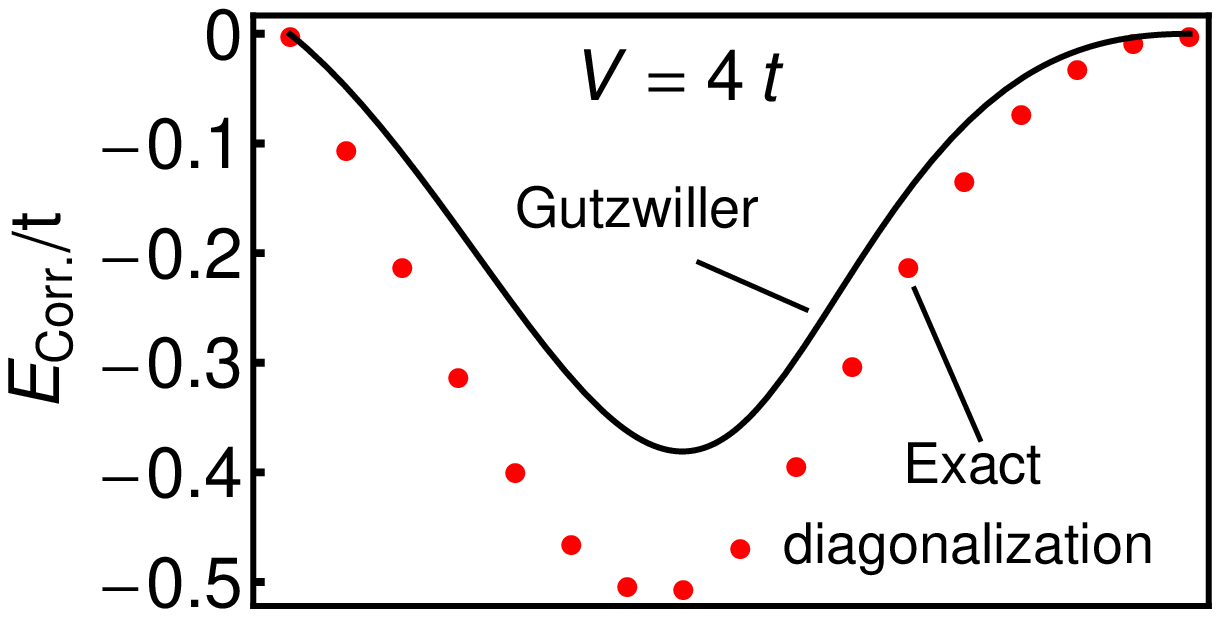}
\includegraphics[angle=0,width=0.7   \linewidth]{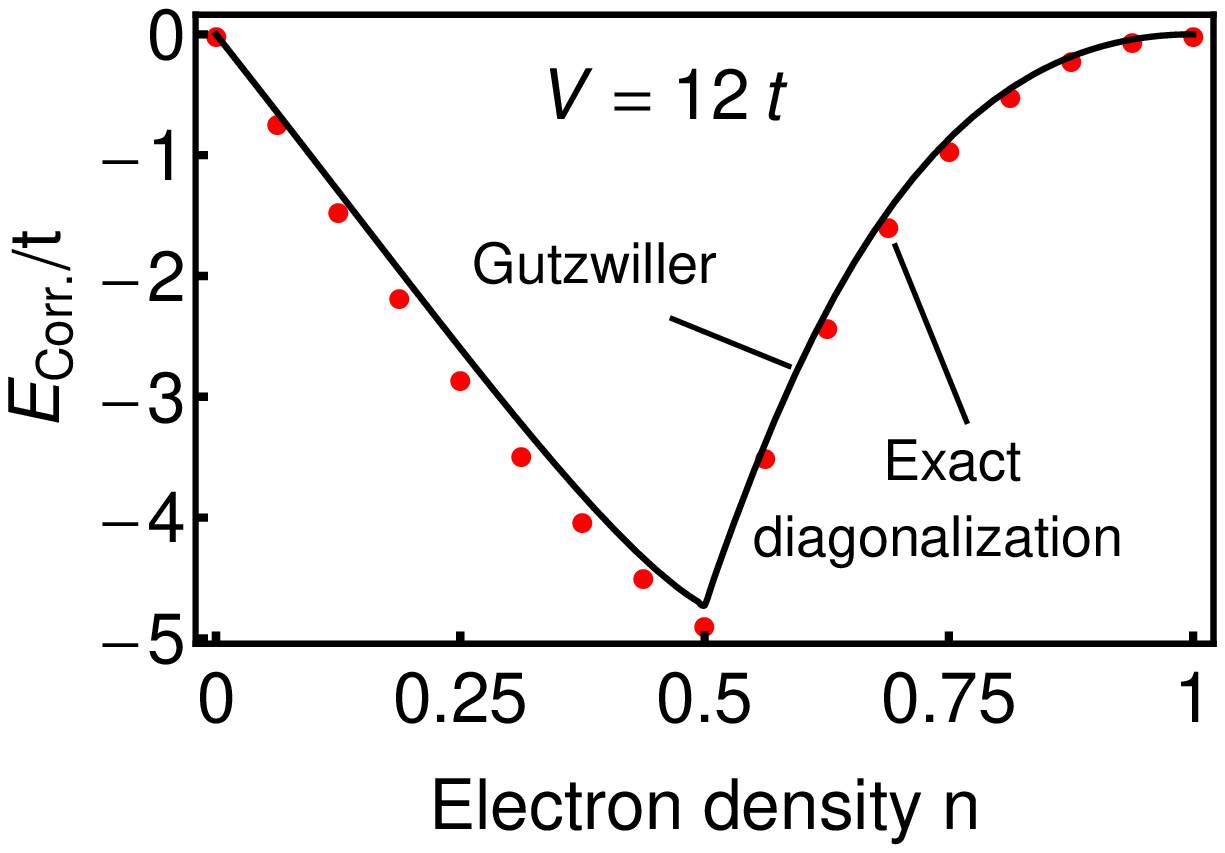}
\caption{(Color online)    Correlation energy per electron obtained from the difference between the ground-state total energy and the same for the uncorrelated state. Correlation energies obtained using the ground-state energy from exact diagonalization (red dots)  and the Gutzwiller approach (solid line) are both shown.
}
\label{Ecorr}
\end{figure}

Now returning to Fig.\ \ref{Eperelectron}, the correlation energy, as expected, is the largest at  half filling, reducing gradually to zero for $n = 0 $ or 1. 
The energy for the uncorrelated state was obtained by evaluating  the kinetic energy for the state
$ |\Psi\rangle   =  \Pi_k^{\text {occ}}    c_k^\dagger |\text {vac} \rangle$, which is simply $E_{ke} = \sum_k^\text {occ} 2t \cos  k$,
 and adding to it the interaction energy $Vb$, where $b =n^2$ is the bond occupancy for the uncorrelated state.  
The correlation energy, defined to be the difference between the exact energy and  the uncorrelated energy, is plotted in Fig.\ \ref{Ecorr}.
It is clear that the Gutzwiller approach is able to account for most of   the correlation energy.

%: Fig 4 

%: Half-filled case
{\it Half-filled case} -- The above expressions are valid for any general filling factor $n$.
We now specialize to the half-filled case, $n = 1/2$, and compare with the differing results of the earlier authors, the resolution of which was in fact one of the motivations for this work. 
By substituting $n=1/2$ into Eqs.\   \eqref{b} and\   \eqref{gamma}, we readily obtain the results for the bond occupancy and the reduction factor
\begin{equation}
b =2^{-1} g (1+g)^{-1} 
\label{bond} 
\end{equation}
and
\begin{equation}
\gamma = 8g (1+g)^{-3}.
\label{gamma-half-filled} 
\end{equation}

The total energy per site in the half-filled case, easily found with the help of Eqs.\   \eqref{EB} and\   \eqref{GTG2}, is  given by   
\begin{equation}
E = - \frac{4g t}{(1+g)^3}  +  \frac{g}{2 (1+g)} V. 
\label{E-half}
\end{equation}
Minimizing the energy with respect to the variational parameter $\partial E/ \partial g = 0$, one finds that at the minimum  
\begin{equation}
g_{0} =  \frac { (v^{-1}+3)^{1/2}  }  {v^{1/2}}    -1 -v^{-1},
\label{g0}
\end{equation}
 if $v \equiv V/(8t) \le 1$, and $g_0 =0$, if $v > 1$. The ground-state energy is plotted in Fig.\   \ref{GS-energy} as a function of the strength of the interaction $V$.

%:4-site model

Note that for the case $V=0$, $g_0 = 1/2$ and the kinetic energy reduction factor becomes $\gamma = 32/27$, i.\ e., greater than one, a result that has already been noted in the literature.\cite{Fazekas2} At first site, this appears counterintuitive because one does not expect the energy of the uncorrelated state to be changed by the Gutzwiller variational wave function when $V=0$, so that $\gamma$ should equal one.
However, this need not be considered surprising since the equal-weight uncorrelated wave function $|\Psi_0\rangle$, used to obtain the reduction factor $\gamma$, does not necessarily have the lowest kinetic energy, even for the case of $V=0$. Therefore the Gutzwiller variational method tries to produce a better ground state and ground-state energy within the variational degree of freedom it is given. The kinetic energy of the correlated state could therefore in principle be larger than the uncorrelated trial state with the kinetic energy reduction factor exceeding one as is the case here.	

We illustrate this with a specific example. We consider   the four-site, two-electron ring model. For this case, one can readily show for $V=0$ that the equal-weight state 
$|\Psi_0\rangle = 6^{-1/2}\sum_{k=1} ^6 | \Gamma_k \rangle$ has the kinetic energy of $-4t/3 $, while the  Gutzwiller variational wave function, 
 $|\Psi_G\rangle = |\Gamma_1 \rangle+ |\Gamma_2 \rangle+ g_0(|\Gamma_3\rangle+|\Gamma_4  \rangle  +|\Gamma_5  \rangle  + |\Gamma_6\rangle  )              $, 
where $g_0 = 1/\sqrt 2$
and the last four configurations have the bond occupancy $B = 1$, while for the first two, $B=0$, 
has a better kinetic energy $-\sqrt 2 t$. Thus the reduction factor, which is given by the ratio of these two energies, becomes 
$\gamma   = (-\sqrt 2 t)/(-4t/3)      \approx 1.06$. The
exact ground-state energy  $-2t$ is much lower. 
The factor $\gamma$ is greater than one, although different from the infinite-lattice value of $\gamma = 32/27$, which it will approach as the number of sites in the ring is increased from four to infinity.
%Thus in the Gutzwiller treatment, the kinetic energy is enhanced by a factor greater than one as compared to the same for the uncorrelated state, because the equal-weight state used as the uncorrelated state   in the Gutzwiller wave function   does not necessarily have the best kinetic energy, even when the interaction $V$ is zero.

%
%: Fig 5
%
%: Exact
{\it Exact solution} -- 
The spinless fermion model in 1D has been exactly solved for the half-filled case  by invoking its equivalence with the XXZ spin Hamiltonian via the Jordan-Wigner transformation.\cite{Giamarchi}  The equivalent Hamiltonian is: $ {\hat H} = {\hat H}_{\text {spin}} + VL/4,$ where
\begin{equation}
{\hat H}_{\text {spin}}=\sum_i J_{xy} ({\hat S}^x_{i+1}{\hat S}^x_i+{\hat S}^y_{i+1}{\hat S}^y_i)+J_z{\hat S}^z_{i+1}{\hat S}^z_i,
\label{eq:XXZ}
\end{equation}
with $J_{xy} = 2t$, $J_z = V$, and $L$ is the number of sites in the 1D lattice. Using its exact solution obtained from the Bethe ansatz,\cite{Bethe}  the  energy for the spinless 1D model, Eq.\   \eqref{Hamil}, may be written as\cite{DesCloizeaux1}
\begin{align}
E =\begin{cases}
\frac{V}{4}- 2t \sin\theta     \displaystyle \int^{\infty}_0 (1-\frac{\tanh \omega \theta}       {\tanh\omega\pi} )     \ d\omega &
 \text{if $0 \leqslant V <2t$ },     \\
2t (  \frac{1}{4}-\ln2)   &         \text{if $V=2t$}, \\
\frac{V}{4}-     2t \sinh\phi\left[\displaystyle \sum^{\infty}_{n=1}(1-\tanh n\phi)+    \frac{1}{2}\right]& \text{if $2t \leqslant V$ },
\end{cases}
\label{exact}
\end{align}
where $\theta \equiv \cos^{-1} (2^{-1} V/t)$ and $\phi \equiv \cosh^{-1} (2^{-1} V/t) $.
The exact energy is  plotted in Fig.\   \ref{GS-energy} together with the Gutzwiller energy, Eqs.\   \eqref{E-half} and\   \eqref{g0}, as well as the results of the slave-boson mean-field theory for comparison. 

%:Slave boson
{\it Slave-boson mean-field solution} -- 
The Gutzwiller reduction factor for the half-filled case, Eq.\   \eqref{gamma-half-filled}, agrees with the results of Fazekas\cite{Fazekas2}, but not with the other two papers in the literature.\cite{Fazekas1,Seibold} The reason for disagreement with the last paper is due to the 
neglect of correlation between the occupancies of the neighboring bonds
in Ref.\  \onlinecite{Seibold}, which in effect turns the intersite interaction problem into an onsite problem. The slave-boson mean-field approach neglects the same correlation as well. In order to understand this point, recall that in the latter method, auxiliary boson fields are introduced corresponding to
the empty, singly, and doubly occupied bonds, which must satisfy certain constraints for the physical Hilbert space. In the mean-field theory, if these constraints are satisfied on each bond, without regard to the state of the neighboring bonds, it leads to the kinetic energy reduction factor in the slave-boson theory
\begin{equation}
\gamma_{\text{sb}} \equiv \langle z z^\dagger \rangle = \frac{(n-b) [\sqrt b+ \sqrt{1-2n +b}]^2} {n (1-n)}, 
\label{gamma2}
\end{equation}
which is the same result for the Hubbard model with on-site interaction, obtained by using either the Gutzwiller or  the slave-boson    treatment of the problem.\cite{Kotliar, Fulde}
 We have checked that even if we keep nearest-neighbor bond correlations within the slave-boson mean-field theory, the results do not change. Furthermore, a similar correlation is missing within the Ogawa\cite{Ogawa} treatment of the Gutzwiller problem, which also produces the same reduction factor Eq.\ (\ref{gamma2}).\cite{Seibold} This is to be compared
with the equivalent reduction factor from our treatment, readily obtained from Eqs.\   \eqref{g2} and\   \eqref{gamma}:  
\begin{equation}
\gamma = \frac{(n-b)^2 [\sqrt b+ \sqrt{1-2n +b}]^2} {n^2 (1-n)^2},
\end{equation}
where we have expressed the reduction factor in terms of the bond occupancy $b$ rather than the Gutzwiller variational parameter $g$.
The ground-state energy per site for the slave-boson case is given by
\begin{equation}
E_{\text{sb}} = \gamma_{\text{sb}}   \epsilon_0 + Vb,
\end{equation}
where $\epsilon_0 $ is the exact uncorrelated ground-state kinetic energy, $\epsilon_0 = -2/ \pi$ at half filling, and the energy has to be minimized with respect to $b$. The results are shown in Fig.\    \ref{GS-energy} together with the exact and the Gutzwiller results. 
%The Gutzwiller energy shown in the figure was computed with the equal-weight state, Eq.\  \eqref{Psi-equal}, as the uncorrelated wave function; A better uncorrelated wave function would provide results closer to the exact energy.

{\it Wigner crystallization} -- Unlike the repulsive Hubbard model in 1D at half-filling, where the ground state is insulating\cite{Lieb} for all non-zero $U/t$, the spinless fermion model does have a metal-insulator transition for the parameter $V/(2t) = 1$. The Gutzwiller approach does give a Brinkman-Rice type metal-insulator transition, although the transition point occurs at $V/(8t) = 1$, which is much higher than the exact result. The transition may be seen from the bond occupancy $b$, which is readily calculated by substituting the minimum value $g_0$ in the expression Eq.\   \eqref{bond}, with the result
$b =6^{-1} (2-\sqrt {1+3V/8t  }).$
%
%\begin{equation}
%      b =\frac{1}{6} (2-(1+\frac{3V}{8t}   )^{1/2}).
%\end{equation}
%
The bond occupancy is a diminishing function of $V$ and becomes zero at and beyond the transition point, leading to the ``Wigner crystal," where alternate sites are occupied.

%:Summary

{\it Summary} -- In Summary, we obtained the Gutzwiller solution of the spinless fermion model in 1D containing the nearest-neighbor Coulomb interaction. Results were presented for a general band filling and compared with the exact solutions. 
It was shown that the Gutzwiller energy compares very well with the exact energy for intermediate coupling. Combinatorials arising in the Gutzwiller solution of the same problem in higher dimensions appear to be quite difficult, but would be a worthwhile problem to study in the future.

This research was supported by the U.S. Department of Energy, Office of Basic Energy Sciences, Division of Materials Sciences and Engineering under Award    No. 
DE-FG02-00ER45818.

%:appendix
\appendix

\section{ Combinatorials}

Here we outline the results for the combinatorials needed for the evaluation of the expectation values for the Gutzwiller wave function for the 1D spinless fermion model with the nearest-neighbor interaction.
The symbols  $K_l(B,N,L)$ and $K(B,N,L)$ represent the number of configurations for the linear chain and the ring arrangements, respectively, where $N$ objects are arranged on a lattice of $L$ sites with $B$ bonds between them (a bond is defined to be present if two contiguous lattice sites are occupied).

\subsection{Enumeration of $K_l$ for the chain arrangement}
First of all, note the fundamental counting result that the number of ways to distribute  $n$ identical objects among $k$ distinct containers with at least one object in each container is ${n-1 \choose k-1}$. Each such distribution can be associated with a string of  $n$  zeros and $k-1$ slashes with each slash inserted between two zeros. The above combinatorial represents the number of ways  $k-1$ positions can be chosen for the slashes out of the $n-1$ positions between the zeros.

Now, we turn to the enumeration of the number of configurations $K_l(B,N,L)$ for the linear chain.
When the occupied positions form a contiguous string (with no intervening unoccupied positions), then $B=N-1$. If the occupied positions are separated (by unoccupied positions) into two contiguous strings, then $B=N-2$. In general, if $C$ is the number of contiguous strings into which the occupied positions are separated, then $B=N-C$.

For a given value of $C$, the number of ways to partition $N$ occupied positions into $C$ contiguous strings is ${N-1 \choose C-1}$. This is the same as ${N-1 \choose N-B-1}={N-1 \choose B}$ (since $C=N-B$).

The number of ways to distribute $C$ contiguous strings among $L-N$ unoccupied positions (so that the contiguous strings are separated from each other by unoccupied positions) is ${L-N+1 \choose C}={L-N+1 \choose N-B}$. (There are $L-N+1$ available locations in which we can insert the $C$ contiguous strings of occupied positions.)
Therefore, the total number of ways to specify $N$ occupied positions among $L$ total positions so that there are exactly  $B$ bonds is given by
\begin{align}
K_l(B,N,L)={N-1\choose B}{L-N+1\choose N-B}.
\label{APeq:NL}
\end{align}
Note that the total number of ways to arrange the $N$ objects on $L$ sites irrespective of the number of bonds is simply: $\sum^{N-1}_{B=0} K_l(B,N,L)={L\choose N} = N_\Gamma$, as it must be, where in the last expression $N_\Gamma$ is the total number of many-body configurations.

\subsection{Enumeration of $K$ for the ring arrangement}

We choose one position on the circle (it doesn't matter which one) and designate it as $p_0$. We now consider two cases.
\begin{enumerate}
\item Position $p_0$ is unoccupied. The number of ways to distribute the $N$ occupied positions among the remaining positions on the circle can be obtained by using the formula obtained for linear arrangements in Eq. \eqref{APeq:NL}. In this case, we have a total of $L-1$ positions. Therefore, the number of admissible arrangements is ${L-1-N+1\choose N-B}={N-1\choose B}{L-N\choose N-B}$.

\item Position $p_0$ is occupied. Start with a circle of $N$ occupied positions, including $p_0$. We need to break the circle into $N-B$ contiguous strings by inserting unoccupied positions among the occupied positions. This means that we need $N-B$ break points. There are $N$ possibilities for break points. Therefore, there are ${N \choose N-B}={N\choose B}$ ways to choose the break points. Having chosen the $N-B$ break points, there are now ${L-N-1\choose N-B-1}$ ways to insert the $L-N$ unoccupied positions. So the total number of circular arrangements with $p_0$ occupied is ${N \choose B}{L-N-1\choose N-B-1}$.
\end{enumerate}

The total number of circular arrangements for a specified value of $B$ is then given by the sum of the above two results. By direct computation, it can be shown that this sum is equal to
\begin{align}
K(B,N,L)=\frac{L}{L-N}{N-1\choose B}{L-N\choose N-B},
 \label{APeq:NCB}
\end{align}
which is the desired result.

\subsection{Enumeration of the number of ring configurations $K_{ij}$ with two adjacent positions specified }

Here, we identify two adjacent positions on the ring and specify in advance whether or not each of those positions is occupied. We designate the number of configurations by $K_{ij}$, where $i, j = 1$ or $0$, where $i$ indicates the occupancy of one site and $j$ indicates the occupancy of the adjacent sites. It clearly does not matter which pair of adjacent sites is occupied from the symmetry of the ring.

\begin{enumerate}
\item The two positions are \emph{both occupied}. The argument is very similar to that of case (2) in the subsection above.
Suppose positions $p_0$ and $p_1$ are occupied. Now, consider a ring of $N$ occupied positions, including $p_0$ and $p_1$. 
We need to break the ring into $N-B$ contiguous strings by inserting unoccupied positions among the occupied positions.
This means that we need $N-B$ {\it break points}. There are  $N-1$ possibilities for the break points (since we can not break between $p_0$ and $p_1$).
So, there are ${N-1\choose N-B} = {N-1\choose B}$ ways to choose the break points. Having chosen the $N-B$ break points, there are now
${L-N-1\choose N-B-1} $ ways to insert the $L-N$ unoccupied positions. Therefore, the total number of  arrangements with both $p_0$ and $p_1$
occupied is
    \begin{align}
   & K_{11}(B,N,L) =
     {N-1\choose B-1}{L-N-1\choose N-B-1}  &  \nonumber  \\  
  &  =\frac{B}{L} K(B,N,L). &
    \label{eq:APN11}
    \end{align}
\item The \emph{first} position is \emph{occupied} and the \emph{second} is not.
Then $N-1$ occupied positions and  $L-N-1$ unoccupied positions remain to be assigned. We need to identify $N-B-1$ break points in the occupied positions and then distribute the unoccupied positions. The number of ways to do this is
    \begin{align}
  &  K_{10}     (B,N,L)={N-1\choose N-B-1}{L-N-1\choose N-B-1}  & \nonumber \\
    &={N-1\choose B}{L-N-1\choose N-B-1}     & \nonumber \\
    &    =\frac{N-B}{L}K(B,N,L). &
    \label{eq:APN10}
    \end{align}

\item The \emph{first} position is \emph{unoccupied} and the \emph{second} is \emph{occupied}. This is obviously the same as case (2), viz.,   $K_{01} (B, N, L) = K_{10} (B, N, L)$, as dictated by symmetry.

\item \emph{Neither} of the two positions is \emph{occupied}. This is similar to case (1) in subsection 2 above. The number of arrangements is
    \begin{align}
    K_{00}(B,N,L)&={N-1\choose B}{L-N-1\choose N-B} \nonumber \\
    &=\frac{L-2N+B}{L} K(B,N,L).
    \label{eq:APN01}
    \end{align}
\end{enumerate}

Note that when we add up these four results, we immediately see that $K_{11} + K_{10} + K_{01} + K_{00} = K$, which is the total number of configurations, irrespective of the occupations of the two chosen adjacent positions.

\subsection{Enumeration of the number of ring configurations $K_{ijkl}$ with four adjacent positions specified }

Extending the notation of the last subsection, $K_{ijkl}$ denotes the number of configurations, where $ijkl$ (each 1 or 0) denote the occupancies of the four consecutive sites. Specifically, we evaluate the quantities $K_{i01l}$, which we will need in the computation of the expectation values for the kinetic energy operator $\langle c_j^\dagger c_k \rangle$. The arguments are analogous to those used in the cases above. The results are:
 \begin{align}
  &  K_{1011}     (B,N,L)={N-2\choose B-1}{L-N-2\choose N-B-2}  & \nonumber \\
    &    =\frac{B(N-B-1)}{(N-1) (L-N-1) }K_{01} (B,N,L), & \\
    & K_{0011}     (B,N,L)={N-2\choose B-1}{L-N-2\choose N-B-1},     & \\
    & K_{1010}     (B,N,L)={N-2\choose B}{L-N-2\choose N-B-2},  &  \\
    & K_{0010}     (B,N,L)={N-2\choose B}{L-N-2\choose N-B-1}.  &
     \end{align}
These four values are denoted as $K_1, K_2, K_3,$ and $K_4$, respectively, in Fig.\   \ref{config} in the main body of the text.
They can also be written as fractions of $K_{01}$ as we have shown explicitly for the first case and these fractions add up to one, so that
$ K_{1011}   
+  K_{0011}   
+ K_{1010}    
+ K_{0010}    = K_{01}  $, as it must since $K_{01}  $ is the total number of configurations without any reference to the occupancies of the first and the fourth sites.

These quantities are difficult to enumerate on the linear chain as the circular symmetry is broken and they become dependent on the positions of the contiguous sites on the linear chain. Because of this, it is easier to evaluate the Gutzwiller energy on the circular ring rather than the linear chain. Both should of course converge to the same result in the thermodynamic limit.

%:References

\end{document}